\documentclass[pre,aps,floats,superscriptaddress,floatfix,twocolumn,longbibliography]{revtex4-1}
\usepackage{amssymb,amsmath}
\usepackage{amsmath,amssymb}
\usepackage{enumitem,kantlipsum}
\usepackage{lipsum}
\usepackage{soul}
\usepackage{cancel}
\usepackage{gensymb}
\usepackage{graphicx}
\usepackage{psfrag}
\usepackage{color}
\usepackage{dcolumn}
\usepackage{bm}
\usepackage[normalem]{ulem}
\usepackage{comment}
\usepackage{hyperref}
\hypersetup{
  colorlinks=true,
  citecolor=magenta,
  linkcolor=cyan,
}

\newcommand{\gap}{\quad \quad \quad \quad \quad}
\begin{document}
\title{Anisotropy can make a moving active fluid membrane rough or crumpled}
\author{Debayan Jana}\email{debayanjana96@gmail.com}
\affiliation{{Theoretical Physics Division}, Saha Institute of Nuclear Physics, a CI of Homi Bhabha National Institute, 1/AF, Bidhannagar, Calcutta 700064, West Bengal, India}
\author{Astik Haldar}\email{astik.haldar@gmail.com}
\affiliation{ Center for Biophysics \& Department of Theoretical Physics, Saarland University, 66123 Saarbr\"ucken, Germany}
\author{Abhik Basu}\email{abhik.123@gmail.com, abhik.basu@saha.ac.in}
\affiliation{{Theoretical Physics Division}, Saha Institute of Nuclear Physics, a CI of Homi Bhabha National Institute, 1/AF, Bidhannagar, Calcutta 700064, West Bengal, India}
\affiliation{Max-Planck-Institut f\"ur Physik komplexer Systeme, N\"othnitzer Strasse 38, 01187 Dresden,Germany}

\begin{abstract}
We present a hydrodynamic theory of anisotropic and inversion-asymmetric moving active permeable fluid membranes. These are described by an anisotropic Kardar-Parisi-Zhang equation. Depending upon the anisotropy parameters, the membrane  is either { effectively isotropic} and algebraically rough with translational short, but orientational long-range order, or unstable, suggestive of membrane crumpling.  

\end{abstract}

\maketitle



The generality of the observations about the dynamical properties of the plasma membranes in live biological cells~\cite{alberts,sriram-rev,poincare,sriram-RMP,flick1,flick2} has prompted scientists to construct minimal physical descriptions. It is, for instance, now understood that the macroscopic membrane properties in mixed lipid membranes depend on whether the membrane is in equilibrium~\cite{tirtha-mem-PRE} or active~\cite{tirtha-mem-NJP}. 
In this article, we set up the hydrodynamic theory of a nearly flat patch of an anisotropic and inversion-asymmetric moving permeable active fluid membrane. The activity could be due to an active species living on the membrane or due to activity in the embedding medium. For simplicity, we assume that there are no conservation laws or broken symmetry modes associated with the activity. Thus, if the activity is due to an active species, e.g., a membrane-spanning active protein, then its concentration is not assumed to be conserved, or if it is due to the activity of the embedding medium, e.g., an orientable active fluid, then it is in its isotropic phase. 
 We  consider an anisotropic membrane.  
 We argue that the, assuming Rouse dynamics~\cite{cai} for simplicity, hydrodynamic equation for the undulation modes of an anisotropic membrane given by a single-valued height field $h({\bf x},t)$ is given by 
  \begin{eqnarray}
{\frac{\partial h}{\partial t}+\frac{\lambda}{2} {({\boldsymbol\nabla}h)^2} = \nu_{ij} \partial_i \partial_jh +f. \label{aniso-kpz}}
\end{eqnarray}
Here $\nu_{ij}$ is symmetric under the interchange of $i$ and $j$ and is an effective anisotropic tension. Noise $f$ has zero-mean and a variance given by
\begin{equation}
{\langle f({\bf x},t)f({\bf 0},0) \rangle = 2D \delta^2({\bf x}) \delta(t). \label{noise}}
\end{equation}
In the absence of any detailed balance, the noise strength $D$ has no specific relation to the damping coefficient matrix $\nu_{ij}$. For arbitrary $\nu_{ij}$, Eq.~(\ref{aniso-kpz}) is an anisotropic generalization of the well-known isotropic Kardar-Parisi-Zhang (KPZ) equation~\cite{stanley}. 
   Since in general $\nu_{ij}\neq 0$ for $i\neq j$, Eq.~(\ref{aniso-kpz}) is anisotropic. We set $\nu_{xx}=\nu_{x}$, $\nu_{yy}=\nu_y$, with $\nu_x\neq \nu_y$ both positive and $\nu_{xy}=\nu_{yx}=\overline\nu$ can be of any sign.{ We point out that even though Eq.~(\ref{aniso-kpz}) has a mathematical form different from the standard anisotropic KPZ form studied previously~\cite{wolf,frey-tauber,john-prx}, a standard coordinate transformation that diagonalizes diffusion tensor can actually bring Eq.~(\ref{aniso-kpz}) to a form same as in Refs.~\cite{wolf,frey-tauber,john-prx}; see Appendix~\ref{app:E}}. See also Ref.~\cite{kardar-new} for detailed studies on the stationary states of anisitropic KPZ equations.
 
\begin{figure*}[]
 \includegraphics[width=\textwidth]{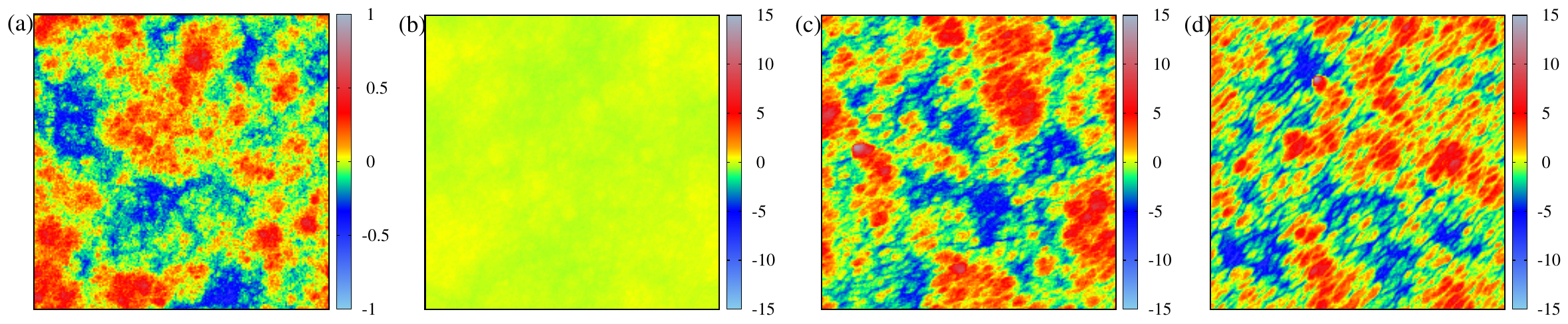}
\caption{ Snapshots of the height profiles (\(L=200\)) in the steady states for (a) and (b) algebraic rough phase, with  parameters (a) \(\nu_x=0.50\), \(\nu_y=0.45\), \(\overline{\nu}=0.10\), \(D=0.005\), and \(\lambda=24\) and (b)  \(\nu_x=0.50\), \(\nu_y=0.55\), \(\overline{\nu}=0.20\), \(D=0.005\), and \(\lambda=24\). Snapshots (c) and (d) correspond to the crumpled phase, with parameters (c) \(\nu_x=1.0\), \(\nu_y=0.20\), \(\overline{\nu}=1.0\), $\nu_4=0.10$, \(D=1.0\), and \(\lambda=2.20\) (with ripples at an arbitrary angle) and (d) \(\nu_x=1.0\), \(\nu_y=1.0\), \(\overline{\nu}=2.0\), $\nu_4=0.10$, \(D=1.0\), and \(\lambda=2.30\) (with ripples at $\pi/4$). Notice the much smaller fluctuations in the algebraic rough phase compared to the crumpled phase (see text).
}\label{density_plot}
\end{figure*}

The most prominent  result from direct numerical solutions (DNS) of Eq.~(\ref{aniso-kpz}) in two dimensions (2D) is that it admits different membrane conformations $h({\bf x},t)$ having strikingly distinct statistical properties controlled by the model parameters. See Fig.~\ref{density_plot} for  representative snapshots of $h({\bf x},t)$ {  falling} in two distinct phases (color bars represent $h$ relative to its  mean) depending on the degree of the surface anisotropy controlled by $\nu_x, \nu_y ,\overline \nu$. The corresponding movies MOV1 [for Fig.~\ref{density_plot}(a)], MOV2 [for Fig.~\ref{density_plot}(c)] and MOV3 [for Fig.~\ref{density_plot}(d)] are available in {Supplemental Material (SM)~\cite{supple}}. We characterize these phases by numerically solving Eq.~(\ref{aniso-kpz}) and  measuring width ${\mathcal W}\equiv \sqrt {\langle [h ({\bf x},t)-\overline h(t)]^2\rangle}$ and by averaging over different realizations, which as a function of time $t$  has an initial growing part, followed by a saturated part ${\cal W}_\text{sat}$ that fluctuates about a steady mean~\cite{stanley}; $\overline h(t)$ is the spatially averaged mean height at time $t$. We find ${\mathcal W}_\text{sat}$ to scale as $L^{\chi}$, respectively, for Fig.~\ref{density_plot}(a); $\chi=0.38$; in our DNS studies, $L$ is system size. See Fig.~\ref{num_sat_region}. Hence we name it algebraic rough phase. For Figs.~\ref{density_plot}(c) and~\ref{density_plot}(d), no steady states are reached  (within the computation time), giving ever increasing fluctuations, suggesting instability and a complete loss of order. Snapshot in Fig.~\ref{density_plot}(b) represent the algebraic rough phase for a different set of model parameters over enhanced ranges as used in Figs.~\ref{density_plot}(c) and~\ref{density_plot}(d) for the crumpled phase, indicate significantly larger fluctuations in the latter. See also later for the detailed numerical results from our DNS studies.
 
 {  We show that our model exhibits algebraic rough phase and is characterized by  translational short-range order (SRO) and orientational long-range order (LRO) and specifically belongs to the KPZ universility class. Using DNS  we verify for different sets of model parameters that the scaling exponents remain almost same indicating the putative algebraic rough phase is model parameter independent and indeed belongs to the KPZ universility class.  Further, the  algebraically rough phase is effectively isotropic with inversion-asymmetry and translational SRO and orientational LRO, akin to a 2D KPZ surface~\cite{kpz,natterman,stanley} and the crumpled phase has translational and orientational SRO, respectively.}  
 
We now derive Eq.~(\ref{aniso-kpz}) and use it to establish the above results and obtain the phase diagrams. The membrane configurations are given by three-dimensional position vector ${\bf R}({\vec s},t)$, where ${\vec s}=(s_1,s_2)$ is a two-dimensional position vector embedded on the membrane, which label points on the membrane. The local membrane velocity ${\bf v}({\vec s},t)\equiv \partial {\bf R}/\partial t$, for a convected or embedded coordinate system~\cite{convec1,convec2}. We now specialize for nearly flat inversion-asymmetric active membrane patches without any overhangs, for which the Monge gauge~\cite{nelson-book} is a convenient choice, in which $(s_1,s_2)\equiv (x,y)$ and ${\bf R}({\bf x})=({\bf x}, h({\bf x})),\, {\bf x}=(x,y)$.  The only hydrodynamic variable is $h({\bf x},t)$, measured with respect to an arbitrary base plane in the Monge gauge~\cite{nelson-book}. 
 Due to the breakdown of inversion symmetry, the patch, considered permeable, is generically moving in a direction local normal to itself.  A change in the membrane conformation  $h({\bf x},t)$ can take place by the drift of the membrane due to activity and relaxational dynamics, which we assume to be Rouse dynamics for simplicity. The general equation of motion of $h({\bf x},t)$, measured with respect to an arbitrary base plane in the Monge gauge~\cite{nelson-book},  reads
 \begin{equation}
  {\frac{\partial h}{\partial t}=v\sqrt{1+ ({\boldsymbol\nabla}h)^2}-\Gamma\frac{\delta {\cal F}}{\delta h}+f,\label{eq_motion}}
 \end{equation}
 Here, ${\cal F}=\int d^2x [\frac{K}{2} (\nabla^2 h)^2 +C\nabla^2 h]$ is a free energy of a tensionless equilibrium fluid membrane (ignoring anharmonic contributions), which governs the relaxation of the membrane to equilibrium in the absence of any activity; here $K>0$ is the bending stiffness~\footnote{Considering for simplicity an isotropic lipid membrane in the absence of any active interactions with the embedding environment, we use an isotropic form of ${\cal F}$, characterized by a single scalar bending stiffness $K$. Since this contribution to the eventual equation of motion turns out to be irrelevant in the long wavelength limit, it leaves our subsequent analysis unaffected.}, ${\boldsymbol\nabla}\equiv (\partial/\partial x,\,\partial/\partial y)$ is the 2D gradient operator, and $C$ is the spontaneous curvature that can be generically present in an inversion-asymmetric membrane and makes the membrane curve one way or the other depending upon its sign. In the Monge gauge, valid for a planar membrane patch, the spontaneous curvature term being a total derivative term vanishes and makes no contribution to $\cal F$, and $\Gamma>0$ is a damping coefficient.  Further, we assume that a small membrane patch moves locally normally to itself at a velocity of magnitude $v$ that is in general a function of the local membrane conformation. Such a normal drift naturally causes $h$ (measured with respect to an arbitrary but fixed reference plane) to change. The square root factor with $v$ on the right-hand side of (\ref{eq_motion}) accounts for the change  in $h$, measured along the $\hat z$ axis due to a $v$ being along the local normal~\cite{stanley}.
  Rotational invariance requires $v$ must remain unchanged under a tilt of the reference plane. This to the linear order in $h$ gives that 
 \begin{equation}
  v(h)\equiv v_0 + \nu_{ij}\partial_i\partial_j h.\label{vperm}
 \end{equation}
In the cellular
context, $v_0$ sets the scale of the drift speed of the membrane, e.g.,
due to actin filament polymerization~\cite{drift} (see also below). The second term on the right-hand side of (\ref{vperm}) gives the contribution from the membrane conformation fluctuations to the drift. Assuming small fluctuations,  we expand the square root in (\ref{eq_motion}) and obtain (\ref{aniso-kpz}),  where the bending energy term scales being $\mathcal{O}(k^4)$ is  irrelevant in the long-wavelength limit and hence is dropped.

A possible microscopic realization of this would be
a nearly flat patch of an inversion-asymmetric,  anisotropic
fluid membrane placed in an isotropic, active suspension of actin
filaments~\cite{sriram-RMP} grafted normally to the membrane locally. Let $\bf p$ be a three-dimensional unit vector  ($p^2=1$) that gives the local orientation of the actin filaments and $\bf n$ be the local unit normal to the membrane; ${\bf n}=(\hat z, -{\boldsymbol\nabla}_i h)$  to the linear order in height fluctuations ($i=x,y$). Locally normal grafting of the filaments implies 
 ${\bf p}\cdot {\bf  n}=1 $, giving $p_i=-\partial_i h$ ($i=x,y$) and $p_z=1$ to the linear order in fluctuations.  The actin filaments exert forces locally normally to the membrane~\cite{chen-perm},  giving a three-dimensional local permeation flow velocity
 \begin{equation}
  {\bf v}\sim v_0 {\bf p} + {\boldsymbol\nabla}\cdot[{\boldsymbol\zeta}{\bf p}{\bf p}]. \label{vp}
 \end{equation}
Now, extracting the component of ${\bf v}$ along the membrane's local normal and to linear order in $h$ yields $v$ as above~\cite{sriram-prl,tirtha-mem-NJP}. Writing the second term on the right-hand side of (\ref{vp}) as the divergence of an active stress ${\boldsymbol\sigma}^a\equiv {\boldsymbol\zeta}{\bf p}{\bf p}$ i.e., $\sigma_{ij}^a\equiv \zeta_{i j \alpha \beta}p_{\alpha}p_{\beta}$, we see that $\nu_{ij}$ is proportional to $\sigma_{ij}^a$.
Tensor ${\boldsymbol\zeta}\equiv\zeta_{ij\alpha\beta}$ encodes the anisotropy in the membrane, e.g., anisotropy in the actomyosin cortex~\cite{acto}, or anisotropic inclusions in the membrane~\cite{inclu}. { Some additional details are given in Appendix~\ref{app:A} for interested readers.} The forms of these anisotropy parameters should depend explicitly on the specific model representation. 
In the absence of anisotropy, in which case the strength of the active stress is described by a scalar parameter, say, $\Delta\mu$, a positive (negative) $\Delta\mu$ refers to extensile (contractile) activity~\cite{sriram-RMP}. Here, however, the active stress is characterized by three parameters, and hence a simple extensile or contractile characterization is no longer possible.


\begin{figure}[t]
\includegraphics[width=0.49\textwidth]{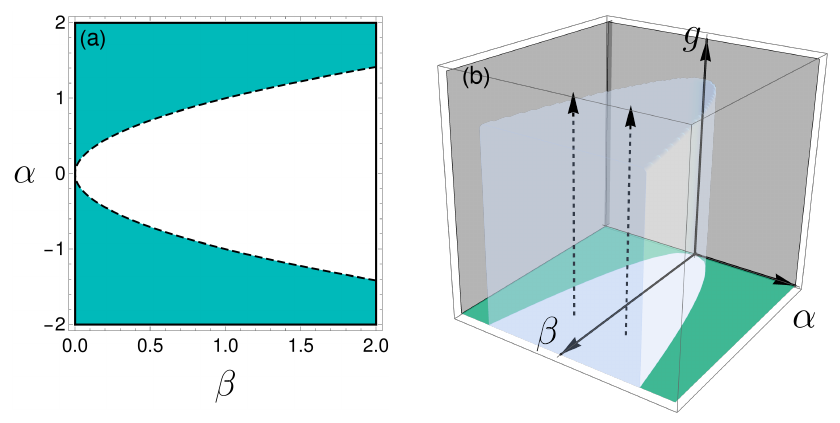}
\caption{ (a) Two distinct phases in $\beta$-$\alpha$ plane. White region is the KPZ like putative algebraic rough phase. (b) The RG flow diagram in the $\beta$-$\alpha$-$g$ space. Arrows pointing away from the $\beta$-$\alpha$ plane represent the direction of RG flows, indicating a strong-coupling algebraic rough phase. Green regions in (a) and (b) correspond to crumpled phase (see text).}
\label{rg_flow_diag}
\end{figure}

 We now theoretically explain the above numerical observations, quantitatively characterizing the phases, by using Eq.~(\ref{aniso-kpz}).
Equation~(\ref{aniso-kpz}) admits a pseudo-Galilean invariance under the transformation $t'=t,\,x_i'=x_i+\lambda u_it,\,h'({\bf x}',t')=h({\bf x},t) + u_i x_i+\frac{\lambda}{2}u^2\,t$ (see Appendix~\ref{app:C}). 
In (\ref{aniso-kpz}), $\nu_{ij}$ plays the role of an (active) anisotropic surface tension. { With $\nu_x, \nu_y>0$, linear stability requires that $\nu_x\nu_y \geq \overline \nu^2$, i.e., $\beta \geq \alpha^2$; otherwise, instability ensues. Here $\beta = \frac{\nu_y}{\nu_x}$ and $\alpha = \frac{\overline \nu}{\nu_x}$. If $\alpha=0$, then Eq.~(\ref{aniso-kpz}) is invariant under (separately) $x\rightarrow -x$ or $y\rightarrow -y$; for $\alpha\neq 0$, the invariance is only under the {\em joint inversion} of $x$ and $y$.
This means $\alpha_c=\pm \sqrt{\beta}$ is a threshold of $\alpha$, with larger $|\alpha|$ implies instability. This is the anisotropy-induced instability mentioned earlier in this article, which can occur even at zero noises, provided $|\alpha|>\sqrt{\beta}$. This then means in general the variance of the local height fluctuations $\langle [h({\bf x},t)-\overline h(t)]^2\rangle$ and local normal fluctuation $\delta {\bf n}= -{\boldsymbol\nabla}h$, $\langle (\delta {\bf n}({\bf x},t))^2\rangle$ diverge.
Can this instability be suppressed by including stabilizing higher-order damping terms, e.g., $- \nu_4\nabla^4 h,\,\nu_4>0$ in (\ref{aniso-kpz})? Such a term can be generically present but is irrelevant if there is no linear instability. In the event of linear instability, these higher-order (in gradient) stabilizing terms are important. Their presence would ensure that  at high-enough wave vectors the system is linearly {stable.} If $\nu_1,\nu_2$ are the eigenvalues of of $\nu_{ij}$, then for membranes of linear size $L< (>) L_c\sim\sqrt{\nu_4/|\text{min}(\nu_1,\nu_2)|}$ are dominated by the stabilizing hyperdiffusion (destabilizing diffusion) terms. In this case, $\langle [h({\bf x},t)-\overline h(t)]^2\rangle$ and $\delta {\bf n}= -{\boldsymbol\nabla}h$, $\langle (\delta {\bf n}({\bf x},t))^2\rangle$ diverge not at every scale, but only if the linear system size $L$ exceeds a threshold that scales with $\sqrt{\nu_4}$~\cite{sm}. Such divergences of $\langle (\delta {\bf n}({\bf x},t))^2\rangle$ for sufficiently larger membranes indicates that such membranes with $L>L_c$ exceeding the instability threshold may be {\em crumpled}, although (\ref{aniso-kpz}) {\em cannot} be used to study the crumpled phase; {see also Appendix~\ref{app:B}} and Ref.~\cite{sm-mbe}. This instability is reminiscent of the strongly crumpled phase in tensionless equilibrium asymmetric tethered membranes~\cite{john-asym-prl,john-asym-pre}. In this unstable phase, the variance of the local normal fluctuations {$\sim \langle ({\boldsymbol\nabla}h)^2\rangle$} diverges for a sufficiently large but finite $L$, giving translational and orientational SRO. In this phase, the growth rate is maximum forming ripples at a particular angle depending on parameters $\nu_x$, $\nu_y$, and $\overline\nu$; see Fig.~\ref{density_plot}(c). Consider the special case where $\nu_x = \nu_y$ and $|\alpha|>1$; then ripples will form at an angle $\pi/4$ ($3\pi/4$) with the $x$ axis for $\alpha>1$ ($\alpha<-1$); see Appendix~\ref{app:B}. See the ripples in Fig.~\ref{density_plot}(d), forming at an angle $\pi/4$, and the corresponding movies MOV2 and MOV3 in {SM~\cite{supple}}.}


{ We now turn to the stable case with $|\alpha|<\sqrt{\beta}$ and consider its stability and fluctuation properties. In the linear theory with $|\alpha|<\sqrt{\beta}$, width ${\mathcal W}_\text{sat} \sim \sqrt {\ln (L/a_0)}$. 
 As in the usual KPZ equation~\cite{kpz}, the combination of noises and nonlinearities here can change the linear theory results substantially. 
{To study the nonlinear effects in the present model~(\ref{aniso-kpz}), it is instructive to perform a one-loop perturbative dynamic
renormalization group (RG) analysis on Eq.~(\ref{aniso-kpz}) with $|\alpha|<\sqrt{\beta}$ and see how the predictions from the RG analysis compare with the results in Ref.~\cite{wolf}.} The Wilson dynamic RG method~\cite{tauber,forster} for our model closely resembles that for the KPZ equation~\cite{kpz,stanley,tauber}. 
Galilean invariance of (\ref{aniso-kpz}) ensures that  $\lambda$ is unrenormalized, which is an exact statement and the analog of the nonrenormalization of the nonlinear coupling in the usual KPZ equation~\cite{stanley,frey-2-loop}; see also Appendix~\ref{app:C}. However, there are diverging one-loop corrections to $\nu_x,\,\,\nu_y,\,\overline\nu$, and $D$, which are represented by the one-loop Feynman diagrams (see Appendix~\ref{app:D}). Dimensional analysis allows us to define an effective dimensionless coupling constant $g$ having a critical dimension 2 as
\begin{equation}
 g=\frac{\lambda^2D}{16\pi(\nu_x\nu_y-\overline\nu^2)^{\frac{3}{2}}},
\end{equation}
along with $\alpha$, $\beta$ defined earlier. The RG differential recursion relations [$\exp(l)$ is a dimensionless length] are 
\begin{align}
   &\frac{dD}{dl}=D\bigg[z-2-2\chi+g\frac{3(1+\beta^2)/4 + \beta/2 + \alpha^2}{(\beta-\alpha^2)} \bigg],\label{D-flow}\\
   &\frac{d\nu_x}{dl}=\nu_x[z-2-g(1/2-\beta/2)]\label{nux-flow},\\
    &\frac{d\nu_y}{dl}=\nu_y[z-2-g(1/2-1/(2\beta))]\label{nuy-flow},\\
    &\frac{d\overline\nu}{dl}=\overline\nu[z-2-g]\label{nubar-flow},\\
    &\frac{dg}{dl}=2g^2\label{g-flow},\\
   &\frac{d\alpha}{dl}=-g(\alpha+\alpha\beta)/2\label{alpha-flow},\\
   &\frac{d\beta}{dl}=g(1-\beta^2)/2\label{beta-flow}.
\end{align}
 Here $z$ and $\chi$ are the dynamic and roughness exponents, respectively~\cite{stanley}.
 Equation~\eqref{g-flow} shows that $g(l)$ diverges as $l\rightarrow l_c\sim {\cal O}(1)$ from below, where $l_c$ is a nonuniversal model parameter-dependent length.  Of course, we cannot follow all the way to $g(l)$ diverging; as soon as $g(l)$ becomes ${\cal O}(1)$, the perturbation theory breaks down and hints a strong-coupling algebraic rough phase similar to the usual 2D KPZ equation~\cite{natterman,stanley}. Assuming $g = g^*$ at the putative perturbatively inaccessible strong-coupling fixed point (SCFP), the flow equations~(\ref{alpha-flow}) and~(\ref{beta-flow}) for $\alpha$ and $\beta$, respectively, yield $\beta = 1$ and $\alpha = 0$ as the {\em stable} fixed points. Thus in the long-wavelength limit ($l \to \infty$), $\nu_x \to \nu_y$ and $\overline{\nu} \to 0$, indicating that the system becomes effectively isotropic and belongs to the KPZ universality class.{ This result can be further rationalized by using the mapping between Eq.~(\ref{aniso-kpz}) and the standard anisotropic KPZ equation studied earlier~\cite{wolf,frey-tauber,john-prx}, together with the 
 well-known result obtained by using a dynamic RG framework in Ref.~\cite{wolf}.  This was  later confirmed by numerical studies in Ref.~\cite{halpin_healy_aniso_kpz}. The flow toward the isotropic strong-coupling fixed point has also been independently demonstrated using nonperturbative RG methods in Ref.~\cite{kloss_aniso}.}

\begin{figure}[b]
\centering
\includegraphics[width=0.49\textwidth]{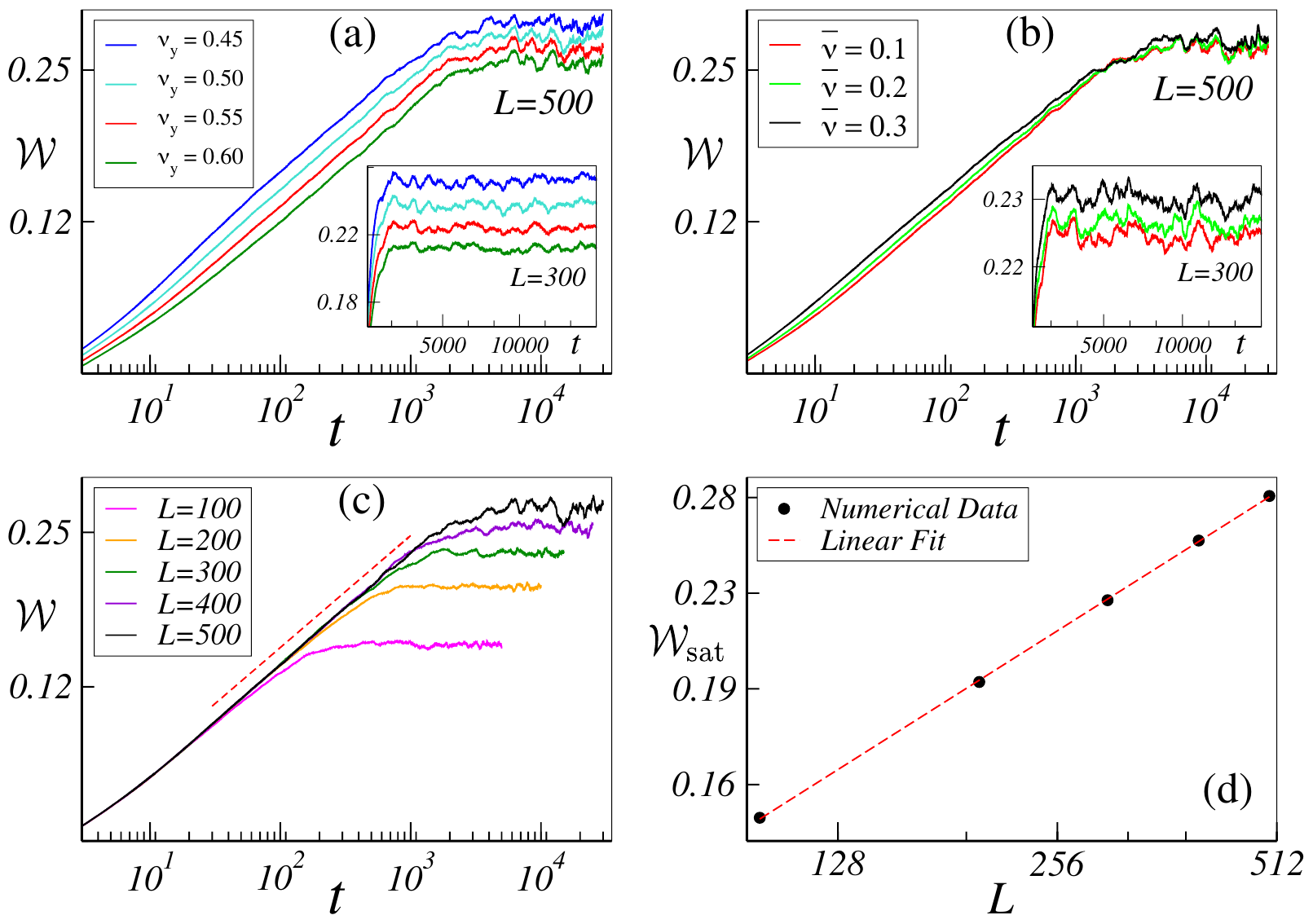}
\caption{{ Log-log plots of $\mathcal{W}$ versus $t$ (a) for different values of $\nu_y$ with fixed $\overline{\nu} = 0.10$ for $L = 500$. Inset: Linear-scale plot for $L = 300$. (b) For varying $\overline{\nu}$ and fixed $\nu_y = 0.55$ with $L = 500$. Inset: Plot for $L = 300$ in linear-scale. (c) For different system sizes with $\nu_y = 0.55$, $\overline{\nu} = 0.20$. The red dashed line indicates a linear fit with slope $\tilde{\beta} = 0.22$. (d) Log-log plot of Saturation width $\mathcal{W}_\text{sat}$ versus $L$ for the same parameters as in (c), with a linear fit yielding $\chi = 0.38$. The parameters $\nu_x = 0.50$, $D = 0.005$, and $\lambda = 24$ are fixed for above plots (see text).}}
\label{num_sat_region}   
\end{figure}

Figure~\ref{rg_flow_diag}(a) shows the phase diagram in the $\beta$-$\alpha$ plane. The green regions at the top and bottom, characterized by $|\alpha| > \sqrt{\beta}$, correspond to the crumpled phase driven by a linear instability. The intervening white region, defined by $|\alpha| < \sqrt{\beta}$, is linearly stable but perturbatively inaccessible within the RG framework. Figure~\ref{rg_flow_diag}(b) displays the corresponding RG flow lines in the full $\beta$-$\alpha$-$g$ space. Flow lines originating within the whitish paraboloid region (bounded by $|\alpha| < \sqrt{\beta}$) escape to a SCFP that is perturbatively inaccessible, reminiscent of the behavior in the 2D isotropic KPZ equation~\cite{stanley,natterman}

To complement the predictions of this RG analysis, we perform DNS of Eq.~(\ref{aniso-kpz}). We discretize the equation using standard forward-backward finite differences on a square grid with lattice spacing $a_0 = 1$ and integrate it numerically via the Euler algorithm with a fixed time step $\Delta t = 0.01$; further implementation details are provided in Appendix~\ref{app:F}. We compute the interface width ${\mathcal W}$ for parameter sets lying in both the (i) speculated algebraically rough regime and (ii) linearly unstable (crumpled) regime of Fig.~\ref{rg_flow_diag}(a).


 Now we focus on the DNS results within the speculated algebraic rough regime, which is argued to be statistically identical to the isotropic 2D KPZ equation. We chose parameter space such that $dg(l)/dl>0$ [see Eq.~(\ref{g-flow})]. Figure~\ref{num_sat_region}(a) shows the $\mathcal{W}(t)$ for various $\nu_y$ values with fixed $\nu_x,\overline\nu$. The linear growth regime exhibits an approximately constant slope across all cases, suggesting that the growth exponent $\tilde{\beta}$ is insensitive to $\nu_y$. However, the $\mathcal{W}_\text{sat}$ increases with decreasing $\nu_y$, as lower $\nu_y$ (for a fixed $\nu_x$) drives the system closer to the instability threshold. See Fig.~\ref{rg_flow_diag}(a). Figure~\ref{num_sat_region}(b) presents $\mathcal{W}(t)$ for different $\overline{\nu}$. Again, the growth exponent remains nearly unchanged, while $\mathcal{W}_\text{sat}$ increases with $|\overline{\nu}|$, consistent with approaching the unstable region at fixed $\nu_x$. See Fig.~\ref{rg_flow_diag}(a). Figure~\ref{num_sat_region}(c) shows variation of ${\cal W}$ over time for different $L$. The red dashed line indicates the slope of the growth region obtained through linear fitting, yielding $\tilde{\beta} = 0.22$. Furthermore, numerically measuring ${\mathcal W}_\text{sat}$ for  $L=100$, $200$, $300$, $400$, and $500$ gives $\chi\approx 0.38$, and hence $z=\chi/\tilde\beta=1.72$ with $\chi+z\approx 2$ (as it should be). See Fig.~\ref{num_sat_region}(d).
Our results on $\chi$ confirms an algebraically rough phase with translational SRO and orientational LRO,  as speculated from our RG calculations above. The value of $\chi=0.38$ compares reasonably with the existing results~\cite{expt_1,expt_2,expt_3,expt_4,expt_5,expt_6,expt_7,expt_8,expt_9,expt_10} and also not far from a mode coupling theory~\cite{jkb-mct} and functional RG~\cite{FRG_1} predictions, indicating 2D KPZ-like scaling behavior lending further credence to a strong-coupling rough phase belonging to the 2D KPZ universality class.

To explore the unstable region, we introduce a regularizing term $-\nu_4 \nabla^4h$ to Eq.~(\ref{aniso-kpz}). For DNS in the unstable region [i.e., green region in Figs.~\ref{rg_flow_diag}(a) and~\ref{rg_flow_diag}(b)] we choose $\nu_4=0.10$, $\Delta t = 0.01$.  However, the interface width still diverges rapidly after a short duration, possibly indicating that additional higher-order nonlinear terms, neglected in (\ref{aniso-kpz}) by using scaling arguments, are now required to describe this region reliably. Figure~\ref{density_plot}(c) [Fig.~\ref{density_plot}(d)], with $L_c = 0.46$ [$L_c = 0.32$], both much smaller than the membrane size $L = 200$, displays the height snapshots just before divergence occurs.{ Such linear instabilities were earlier  pointed out in the context of vicinal surfaces; see footnote [13] of Ref.~\cite{wolf}).}

So far we have considered a nearly flat patch within a Monge parametrization, which holds up to length scales at which curvature $C$ becomes important.  At larger scales, when $C$ could be important, the spontaneous curvature term  scales as $C/r$.  In the algebraic rough phase, $g$ should not vanish, with inversion asymmetry  surviving in the long-wavelength limit. Nonetheless, with expected $z<2$, $\nu_{ij}$ continues to dominate over $C$ (in a scaling sense), ultimately giving a statistically flat but inversion-asymmetric membrane. For the crumpled phase, it is not possible to comment on the large-scale structure of the membrane.



We have thus shown that the hydrodynamic theory for a moving active anisotropic fluid membrane is given by an anisotropic KPZ equation, giving 
an algebraically rough phase, that {\em in spite of the anisotropy} belongs to the isotropic 2D KPZ universality class (and hence isotropy is an emergent symmetry of the model in the long-wavelength limit).{ The prediction of an effectively isotropic algebraically rough phase in the linearly stable sector of Eq.~(\ref{aniso-kpz}) is a direct consequence of its mapping to the anisotropic KPZ equation studied in Refs.~\cite{wolf,frey-tauber,john-prx}, where the large-scale irrelevance of anisotropy has been established. See Appendix~\ref{app:E} for related technical discussions.}
{In addition, our analytical and numerical studies identify an unstable regime, which cannot be controlled by adding higher-order linear damping terms, and the fluctuations continue to grow unbounded in our DNS studies. We thus speculate that this unstable regime actually corresponds to a crumpled phase.} { In calling the unstable phase crumpled, we were guided by the divergence of $\langle h^2({\bf r},t)\rangle$ as soon as the linear system size exceeds a finite threshold.  This criterion is similar to what was used is Refs.~\cite{sm,sm-mbe,john-asym-prl,john-asym-pre}. Nonetheless, we caution the readers that the precise nature of this phase can be of various types, e.g., patterned states, or some symmetry-broken states, which cannot be inferred from our theory.} Our results show that the fluctuation-induced average membrane velocity $\frac{\lambda}{2}\langle (\nabla h)^2\rangle$, proportional to $g$ is finite in the algebraically rough phase. We are unable to comment on it in the unstable phase. In living cells, lamellipodia where actin filaments grow nearly normal to the membrane, realize our model via activity-driven permeation flows~\cite{expt_rev_1,expt_rev_2}. Similarly, {\em in vitro} supported bilayers with actin nucleators exhibit ATP-driven fluctuations and forces~\cite{expt_rev_3} are applicable to our model. Our results suggest that coarse-grained experiments, designed to explore long-wavelength scaling properties, cannot detect microscopic anisotropy when the surface is stable.

{We have considered the embedding medium to be isotropic. This study can be generalized by considering an embedding nematic medium. How extensile or contractile activities affect the ensuing steady states of the membrane should be an important future research direction~\cite{ordered_1, ordered_2, isotropic_1, isotropic_2}.} It would be interesting to explore connections between our predicted unstable phase and observed reduction of membrane tension to even negative values during membrane fusion~\cite{nega-tension}. Membrane tension may be measured using optical setups~\cite{optics}, giving information about $\nu_x,\nu_y$, and $\overline\nu$. Comparison with experiments on membrane fluctuations by using, e.g., spectroscopy~\cite{timo} can give $\lambda$. Studying how the fluid flows along the membrane in the different phases and their dependence on the tension anisotropy are related to the cortical flows reported in Ref.~\cite{acto} and how the membrane dynamics may couple with a diffusing density living on the membrane~\cite{astik-xy-1,astik-xy-2} should be promising future work directions. {One may also consider in which the fluid membrane is tensed, i.e., has a finite surface tension, even in equilibrium. Such a surface tension, isotropic or anisotropic, can be absorbed in $\nu_{ij}$, leaving the form of Eq.~\eqref{aniso-kpz} unchanged and subsequent analysis unaffected.}
It will be interesting to connect Eq.~\eqref{aniso-kpz} to variants of the Toom model~\cite{toom1,toom2} and generalize our study by considering the interplay between the nonlocal effects~\cite{nonlocal-kpz1,nonlocal-kpz2}  and anisotropy.



{\em Acknowledgments.} A.H. thanks Alexander von Humboldt (AvH) Stiftung (Germany) for a postdoctoral fellowship. A.B. thanks the AvH Stiftung (Germany) for partial financial support under the Research Group Linkage Programme scheme (2024). A.B. thanks ANRF (India) for partial financial support through the ARG (MATRICS) programme (File No. ANRF/ARGM/2025/000461/TS).


\appendix

\section{Anisotropic KPZ equation}

{ 
We start with Eq.~(\ref{vp}) of the main text, which 
gives the three-dimensional local permeation flow velocity
 \begin{equation}
  {\bf v}\sim v_0 {\bf p} + {\boldsymbol\nabla}\cdot {\boldsymbol \sigma}, 
 \end{equation}
 which is understood to be in the direction of the local normal to a membrane patch, whose magnitude can be obtained by dotting the above equation with ${\bf n}$, the unit local normal to the membrane patch. Here the active stress tensor ${\boldsymbol \sigma}$ is defined as 
\begin{equation}
{\boldsymbol\sigma} \equiv {\boldsymbol\zeta}\mathbf{p}\mathbf{p}.
\end{equation}


The local orientation of actin filaments is described by a three-dimensional unit vector $\mathbf{p}$, which is assumed to be locally normal to the membrane, i.e., along the unit normal $\mathbf{n}$ to the membrane patch in the Monge gauge,  given by

\begin{equation}
{\mathbf{p} = \mathbf{n} =
\frac{1}{\sqrt{1 + (\boldsymbol\nabla h)^2}}
\left(\hat z, -\frac{\partial h}{\partial x},
       -\frac{\partial h}{\partial y}\right).}
\end{equation}
Thus for a (assumed) nearly flat surface patch, the leading-order departures to the components of $\mathbf{p}$ for a perfectly flat surface due to (small) fluctuations in $h$ are ${\cal O}({\boldsymbol\nabla}h)$ along the $x$ and $y$ directions. 

Now the velocity along the $z$ direction due to this active stress is given by [excluding a possible constant contribution marked by $v_0\mathbf{p}$ in Eq.~(\ref{vp}) of the main text]
\begin{equation}
v_z^{a} = \partial_j \sigma_{z j}^a
= \zeta_{z j \gamma \delta} \, \partial_j (p_\gamma p_\delta).
\end{equation}
Expanding,
\begin{align}
v_z^{a}
&=
\zeta_{zjxx} \, \partial_j (p_x^2)
+ \zeta_{zjyy} \, \partial_j (p_y^2)
+ \zeta_{zjzz} \, \partial_j (p_z^2)
\nonumber \\
&\quad
+ \zeta_{zjxy} \, \partial_j (p_x p_y)
+ \zeta_{zjyx} \, \partial_j (p_y p_x)
\nonumber \\
&\quad
+ \zeta_{zjxz} \, \partial_j (p_x p_z)
+ \zeta_{zjzx} \, \partial_j (p_z p_x)
\nonumber \\
&\quad
+ \zeta_{zjyz} \, \partial_j (p_y p_z)
+ \zeta_{zjzy} \, \partial_j (p_z p_y).
\end{align}
Since $\partial_j (p_z^2)=0$ to the lowest order in $h$, and the terms involving higher powers are of order $\mathcal{O}(k^3)$ and hence irrelevant in long-wavelength limit, only last four terms contribute at leading order to $v_z^a$. Thus  keeping only leading-order terms in the Monge gauge
\begin{align}
v_z^{a}
&=
- \zeta_{zxxz} \frac{\partial^2 h}{\partial x^2}
- \zeta_{zxzx} \frac{\partial^2 h}{\partial x^2}
- \zeta_{zxyz} \frac{\partial^2 h}{\partial x \partial y}
- \zeta_{zxzy} \frac{\partial^2 h}{\partial x \partial y}
\nonumber \\
&\quad
- \zeta_{zyxz} \frac{\partial^2 h}{\partial x \partial y}
- \zeta_{zyzx} \frac{\partial^2 h}{\partial x \partial y}
- \zeta_{zyyz} \frac{\partial^2 h}{\partial y^2}
- \zeta_{zyzy} \frac{\partial^2 h}{\partial y^2}.
\end{align}
In obtaining the above results, any differences between an embedded coordinate defined on the nearly flat patch along the $xy$ plane and the laboratory frame is ignored, since the corrections to the above results from the differences are higher order in $h$. Combining the above results with the constant contribution gives Eq.~(\ref{vperm}) of the main text.
%
%

%
}\label{app:A}

\section{Linear stability analysis}
 We start from Eq.~(\ref{aniso-kpz}) of the main text for an anisotropic membrane with a fixed background, ignoring the conservation of embedding bulk fluid momentum. As argued in the main text, the hydrodynamic equation governing the undulation modes of such a membrane, described by the height field $h({\bf x},t)$, is an anisotropic KPZ equation with a Gaussian-distributed, zero-mean white noise $f({\bf x},t)$
{\begin{eqnarray}
\frac{\partial h}{\partial t}+\frac{\lambda}{2} {({\boldsymbol\nabla}h)^2} = \nu_{ij} \partial_i \partial_jh +f. \label{aniso-kpz-supple}
\end{eqnarray}}

{ Here both $i$ and $j$ can take value $1$ or $2$.}
To perform linear stability analysis, we consider only the linear terms  in Eq.~(\ref{aniso-kpz}), i.e., set $\lambda=0$.  The resulting linearized equation of $h$ in Fourier space is
\begin{align}
 \partial_t h= -\nu_x(k_x^2+\beta k_y^2+ 2\alpha k_x k_y) h,\label{lin-h-f} 
\end{align}
where ${\bf k}=(k_x,k_y)$ is a wavevector. We define
$\beta\equiv \nu_y /\nu_x$, $\alpha\equiv{\overline{\nu}}/{\nu_x}$; see also  the main text. To proceed further, we use the polar coordinates and introduce $k_x=k\cos\theta$ and $k_y=k\sin\theta$, where $k=|{\bf k}|$ and $\theta=\tan^{-1}(k_y/k_x)$ is the polar angle. Now for linear stability, $-\nu_xk^2(\cos^2\theta+\beta\sin^2\theta+\alpha\sin 2\theta)<0$ for all $\theta$. With $\nu_x,\,\nu_y >0$; as assumed in the main text, we must have
$(1+\beta\tan^2\theta+2\alpha\tan\theta)> 0 $ for all $\theta$ to ensure linear stability. This means $\beta \geq \alpha^2$. When \(1 + \beta \tan^2 \theta + 2\alpha \tan \theta < 0\), linear instability occurs. The growth rate of the fluctuations is maximized when \(1 + \beta \tan^2 \theta + 2\alpha \tan \theta\) reaches its most negative value, which occurs at a specific \(\theta\) determined by the values of \(\alpha\) and \(\beta\). Consequently, a stripelike pattern emerges based on the value of \(\theta\). A representative snapshot of this case is shown in Fig.~\ref{density_plot}(c) of the main text. Now we consider the special case $\nu_x=\nu_y=\nu$, i.e., $\beta=1$ and $\alpha=\overline\nu/\nu$. For linear stability, $-\nu k^2(1+\alpha\sin 2\theta)<0$ for all $\theta$. With $\nu>0$, we must have
$(1+\alpha\sin2\theta)> 0 $ for all $\theta$ to ensure linear stability. Since the maximum of $|\sin\,2\theta|$ is unity, the previous condition in turn means $|\alpha|\le 1$, or $\nu\pm\overline{\nu}>0$. When \( (1+\alpha\sin 2\theta) < 0 \), linear instability again ensues. The growth rate of the fluctuations is maximized when \( (1+\alpha\sin 2\theta) \) attains its largest negative value. For \( \alpha > 1 \), this happens at \( \theta = 3\pi/4 \), corresponding to \( k_x = -k_y \) giving the Fourier wavevector with the maximum growth rate. Keeping only the corresponding Fourier mode (which is the most dominant Fourier mode), we get 
\begin{equation}
h(\mathbf{r}) \approx h_{\mathbf{k}} e^{i\frac{k}{\sqrt{2}}(-x+y)},
\end{equation}
which has a phase given by 
\begin{equation}
\phi(x,y) = \frac{k}{\sqrt{2}}(y-x).
\end{equation}
If we move in the \(xy\) plane along paths with a constant phase, i.e., with constant \( y-x \), the exponential factor \( e^{i\phi} \) does not change. Consequently, the height function \( h(x,y) \) does not vary along these directions. These lines, defined by \( y-x = \text{const} \), correspond to the stripes observed in real space, forming at an angle of \( \pi/4 \). The periodicity of the stripes is controlled by the wavenumber \( k_\text{max} \sim \pi/a_0 \sim \pi \), given that \( a_0 = 1 \) in our simulations. A representative snapshot of this pattern can be seen in Fig.~\ref{density_plot}(d) of the main text.
On the other hand, for \( \alpha < -1 \), the maximum growth rate occurs at \( \theta = \pi/4 \), corresponding to \( k_x = k_y \). Now using the logic used above, we can say this corresponds to the formation of stripes at an angle of \( 3\pi/4 \) in the \( x \)-\( y \) plane. { We have verified this result (not shown).}\label{app:B}

\section{Galilean invariance}
We now explicitly show that under the Galilean transformation  defined by
\begin{align}
&t\rightarrow t'\equiv t,\,x_i \rightarrow x_i'\equiv x_i+\lambda u_it,\, h({\bf x},t) \rightarrow \nonumber \\ 
&h'({\bf x}',t') \equiv h({\bf x},t) + u_i x_i +\frac{\lambda}{2}u^2t.
\end{align}
 Equation~(\ref{aniso-kpz}) remains invariant. Under the Galilean transformation
\begin{align}
& \frac{\partial}{\partial t} \rightarrow \frac{\partial}{\partial t'}+\lambda u_i\frac{\partial}{\partial x_i'},\, \frac{\partial}{\partial x_i} \rightarrow \frac{\partial}{\partial x_i'} ,\nonumber\\
& \frac{\partial h({\bf x},t)}{\partial t} \rightarrow \frac{\partial h'({\bf x'},t')}{\partial t'} +\lambda u_i\frac{\partial h'({\bf x'},t')}{\partial x_i'} -\frac{\lambda}{2}u^2. \label{GI1} \\
& \frac{\lambda}{2}\frac{\partial h({\bf x},t)}{\partial x_i}\frac{\partial h({\bf x},t)}{\partial x_i}\rightarrow \frac{\lambda}{2}\frac{\partial h'({\bf x'},t')}{\partial x'_i}\frac{\partial h'({\bf x'},t')}{\partial x'_i}\nonumber \\ &\gap\quad\quad\quad\quad+\frac{\lambda}{2}u^2 -\lambda u_i\frac{\partial h'({\bf x'},t')}{\partial x'_i}. \label{GI2}\\ & \nu_{ij}\frac{\partial}{\partial x_i}\frac{\partial h({\bf x},t)}{\partial x_j}\rightarrow \nu_{ij}\frac{\partial}{\partial x'_i}\frac{\partial h'({\bf x'},t')}{\partial x'_j}. \label{GI3} 
\end{align}
The variance of the noise $f$, being $\delta$-correlated in time, remains invariant under this transformation. Using Eqs.~(\ref{GI1}), (\ref{GI2}), and (\ref{GI3}) and rewriting Eq.~(\ref{aniso-kpz}) in terms of $h'({\bf x'},t')$, ${\bf x'}$, and $t'$, we find that Eq.~(\ref{aniso-kpz}) is Galilean invariant.
\label{app:C}

\section{Renormalization group calculations} 
{

We start by constructing a generating functional~\cite{bausch,tauber} by using Eq.~(\ref{aniso-kpz}) and Eq.~(\ref{noise}) of the main text. We get
\begin{equation}
 \mathcal{Z}=\int \mathcal{D}\hat{h} \mathcal{D}h e^{-\mathcal{S}[\hat{h},h]},
\end{equation}
where $\hat{h}$ is referred as response field and $\mathcal{S}$ is action functional, which is given by
\begin{align} 
{\mathcal{S}= -\int_{x,t}\hat{h}D\hat{h} + \int_{x,t}\hat{h}(\partial_th+\frac{\lambda}{2} {({\boldsymbol\nabla}h)^2}-\nu_{ij} \partial_i \partial_jh).\label{action}}
\end{align}


\subsection{Results from the linearized equation of motion}

This can be obtained from the Gaussian part of the action functional~(\ref{action}), i.e., by setting $\lambda=0$ in (\ref{action}).

We first define Fourier transforms in space and time by
\begin{equation*}
h({\bf x},t)=\int_{{\bf k},\omega}h({\bf k},\omega)e^{i({\bf k\cdot x}-\omega t)}.
\end{equation*}
Then the (bare) two-point functions in the Gaussian limit of (\ref{action}) are 
\begin{subequations}
\begin{align}
&\langle \hat{h}({\bf k},\omega) \hat{h}(-{\bf k},-\omega)\rangle_0=0,\\
&\langle \hat{h}({\bf k},\omega) h(-{\bf k},-\omega)\rangle_0=\frac{1}{i\omega +\nu_{ij}k_ik_j},\\
&\langle \hat{h}(-{\bf k},-\omega) h({\bf k},\omega)\rangle_0=\frac{1}{-i\omega +\nu_{ij}k_ik_j},\\
&\langle h({\bf k},\omega) h(-{\bf k},-\omega)\rangle_0=\frac{2D}{\omega^2 +(\nu_{ij}k_ik_j)^2}.
\end{align}
\end{subequations}

\begin{figure}[h]
\centering
\includegraphics[width=0.2\textwidth]{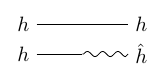}
\caption{Diagrammatic representations of two point functions.}
\label{propagator}
\end{figure}
Figure~\ref{propagator} shows the diagrammatic representations of the various  two-point functions.



\subsection{One-loop corrections to the model parameters}

In the (more usual) isotropic models, applications of the Wilson momentum space RG involves integrating the wave-vector magnitude $q$ within a shell $\Lambda/b \leq q \leq \Lambda,\,b>1$~\cite{halpin,tauber}. The present model is, however, anisotropic, lacking any spherical symmetry. Thus, using spherical coordinates and a spherical shell for mode elimination is not convenient for the present problem. Instead, we integrate $q_x$ over the whole range from $-\infty$ to $\infty$ and then integrate $q_y$ over two thin strips  $\Lambda/b \leq q_y \leq \Lambda$ and $-\Lambda \leq q_y \leq -\Lambda/b$ with $b>1$; see Fig.~\ref{int_diag}. Of course, we could have chosen another  integration scheme, in which $q_y$ is integrated  over the whole range from $-\infty$ to $\infty$, and then integrate $q_x$ over two thin strips  $\Lambda/b \leq q_x \leq \Lambda$ and $-\Lambda \leq q_x \leq -\Lambda/b$. For our model, the two schemes are equivalent, connected to the fact that the different anisotropic terms of the same order scale in the same way. 

Notice that the limits of the $q_x$ integration is  extended from $-\infty$ to $+\infty$, although there is a physical upper wave-vector cutoff$~\Lambda$ in both the $x$ and $y$ directions. The extension of the limits of the $q_x$ integrals {is} done for reasons of calculational convenience and has no effect on the physical results in the long wave-length limit. This is because the integrands vanish in the limit of $q_x\rightarrow \pm \infty$ and the dominant contributions to all the integrals come from small $q_x\sim q_y$. See also Ref.~\cite{john-aniso}.

\begin{figure}[h]
\centering
\includegraphics[width=0.5\textwidth]{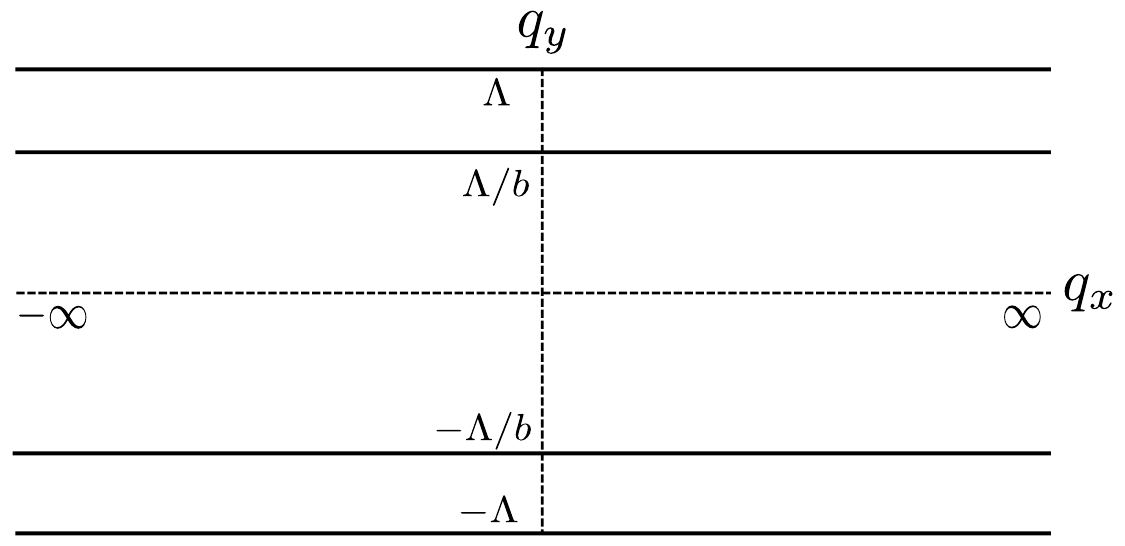}
\caption{Wave-vector-shell integration in the one-loop diagrams obtained by perturbative expansions of the action functional~(\ref{action}), corresponding to the anisotropic KPZ equation. Here $q_x$ is integrated over the range $-\infty$ to $\infty$, while $q_y$ is integrated over two thin strips  $\Lambda/b \leq q_y \leq \Lambda$ and $-\Lambda \leq q_y \leq -\Lambda/b$.}
\label{int_diag}
\end{figure}

We now proceed with the evaluation of the integrals that arise from one-loop Feynman diagrams. These integrals take the form
\begin{align}
    &\int \frac{dq_x}{2\pi} \int \frac{dq_y}{2\pi} \frac{\tilde{f}(q_x,q_y)}{(\nu_x q_x^2 + 2\overline{\nu} q_x q_y + \nu_y q_y^2)^m}\nonumber \\
    &= \frac{1}{\nu_x^m} \int \frac{dq_x}{2\pi} \int \frac{dq_y}{2\pi} \frac{\tilde{f}(q_x,q_y)}{(q_x^2 + 2\alpha q_x q_y + \beta q_y^2)^m} \nonumber \\
    &= \frac{1}{\nu_x^m} \int \frac{dq_x}{2\pi} \int \frac{dq_y}{2\pi} \frac{\tilde{f}(q_x,q_y)}{(q_x - a_+)^m (q_x - a_-)^m}, \label{strip_int_1}
\end{align}
where $\tilde{f}(q_x,q_y)$ represents functions of $q_x$ and $q_y$, and we use $\alpha = {\overline{\nu}}/{\nu}$, $\beta = \nu_y/\nu_x$. The exponent $m$ takes values $2$ or $3$. The terms $a_+ = q_y (-\alpha + i p)$ and $a_- = q_y (-\alpha - i p)$ are two poles of order $m$, where $p = \sqrt{\beta - \alpha^2}$. 

To evaluate this integral, we first perform the integration over $q_x$ in the range $(-\infty, \infty)$. Subsequently, we integrate over the $y$ component of momentum within the two high-$|q_y|$ strips, specifically in the regions $\Lambda/b \leq q_y \leq \Lambda$ and $-\Lambda \leq q_y \leq -\Lambda/b$. See Fig.~\ref{int_diag}. For computational convenience, we define $q_x' = q_x + \alpha q_y$. Under this change of variables, Eq.~(\ref{strip_int_1}) transforms into
\begin{align}
    &\frac{1}{\nu_x^m}\int_{\Lambda/b}^{\Lambda} \frac{dq_y}{2\pi} \int_{-\infty}^{\infty} \frac{dq_x'}{2\pi} \frac{\tilde{f}(q_x',q_y)}{(q_x'^2 + p^2 q_y^2)^m}\nonumber \\ 
    &\quad+ \frac{1}{\nu_x^m}\int_{-\Lambda}^{-\Lambda/b} \frac{dq_y}{2\pi} \int_{-\infty}^{\infty} \frac{dq_x'}{2\pi} \frac{\tilde{f}(q_x',q_y)}{(q_x'^2 + p^2 q_y^2)^m}. \label{strip_int_2}
\end{align}
The first integral receives contributions from the positive values of $q_y$, while the second integral accounts for the negative values of $q_y$. Since the denominator of each of the integrals
is even in $q_x'$,  any terms in $\tilde{f}(q_x',q_y)$ containing odd powers of $q_x'$ must vanish under integration over $q_x'$.  Only terms with even powers of $q_x'$ survive and contribute to the integral. It turns out from the explicit expressions of the various one-loop integrals that these are also even powers of $q_y$. Consequently, both integrals in Eq.~(\ref{strip_int_2}) yield the same result. Hence, in the following analysis, we evaluate the integrals considering only the positive values of $q_y$, noting that the contributions from negative $q_y$ are identical. Below, we list the integrals arising in the calculations and evaluate the one-loop diagrams that account for fluctuation-induced corrections to the model parameters. Here $\int_{q_x,q_y} \equiv \int_{\Lambda/b}^{\Lambda} \frac{dq_y}{2\pi} \int_{-\infty}^{\infty} \frac{dq_x}{2\pi}$.

\begin{widetext}
\begin{align}
&I_1= \int_{q_x,q_y} \frac{q_x^2}{(\nu_x q_x^2 + 2\overline{\nu} q_x q_y + \nu_y q_y^2)^2} = \frac{1}{4\nu_x^2}\frac{\beta}{p^3} \int_{\Lambda/b}^{\Lambda} \frac{dq_y}{2\pi q_y},\nonumber\\
&I_2= \int_{q_x,q_y} \frac{q_y^2}{(\nu_x q_x^2 + 2\overline{\nu} q_x q_y + \nu_y q_y^2)^2} = \frac{1}{4\nu_x^2}\frac{1}{p^3} \int_{\Lambda/b}^{\Lambda} \frac{dq_y}{2\pi q_y},\nonumber\\
&I_3= \int_{q_x,q_y} \frac{q_xq_y}{(\nu_x q_x^2 + 2\overline{\nu} q_x q_y + \nu_y q_y^2)^2} = -\frac{1}{4\nu_x^2}\frac{\alpha}{p^3} \int_{\Lambda/b}^{\Lambda} \frac{dq_y}{2\pi q_y},\nonumber\\
&I_4= \int_{q_x,q_y} \frac{q_x^4}{(\nu_x q_x^2 + 2\overline{\nu} q_x q_y + \nu_y q_y^2)^3} = \frac{3}{16\nu_x^3}\frac{\beta^2}{p^5} \int_{\Lambda/b}^{\Lambda} \frac{dq_y}{2\pi q_y},\nonumber\\
&I_5= \int_{q_x,q_y} \frac{q_x^2 q_y^2}{(\nu_x q_x^2 + 2\overline{\nu} q_x q_y + \nu_y q_y^2)^3} = \frac{3}{16\nu_x^3}\frac{\beta+2\alpha^2}{3p^5} \int_{\Lambda/b}^{\Lambda} \frac{dq_y}{2\pi q_y},\nonumber\\
&I_6= \int_{q_x,q_y} \frac{q_x^3 q_y}{(\nu_x q_x^2 + 2\overline{\nu} q_x q_y + \nu_y q_y^2)^3} = -\frac{3}{16\nu_x^3}\frac{\alpha\beta}{p^5} \int_{\Lambda/b}^{\Lambda} \frac{dq_y}{2\pi q_y},\nonumber\\
&I_7= \int_{q_x,q_y} \frac{q_x q_y^3}{(\nu_x q_x^2 + 2\overline{\nu} q_x q_y + \nu_y q_y^2)^3} = -\frac{3}{16\nu_x^3}\frac{\alpha}{p^5} \int_{\Lambda/b}^{\Lambda} \frac{dq_y}{2\pi q_y},\nonumber\\
&I_8= \int_{q_x,q_y} \frac{q_y^4}{(\nu_x q_x^2 + 2\overline{\nu} q_x q_y + \nu_y q_y^2)^3} = \frac{3}{16\nu_x^3}\frac{1}{p^5} \int_{\Lambda/b}^{\Lambda} \frac{dq_y}{2\pi q_y}.\nonumber
\end{align}
\end{widetext}

\subsection{One-loop corrections to $D$}

The one-loop diagram correcting $D$ is given in Fig.~\ref{D_diag}, which comes with a symmetry factor of $2$. 
\begin{figure}[b]
\centering
\includegraphics[width=0.4\textwidth]{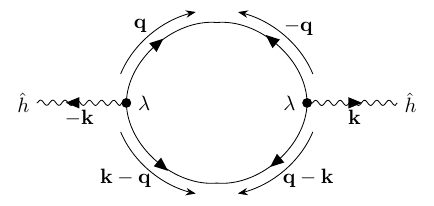}
\caption{One-loop Feynman diagram that contributes to the correction of $D$.}
\label{D_diag}
\end{figure}
It is given by
\begin{align}
\frac{\lambda^2}{4} \int \frac{d\Omega}{2\pi} \frac{d^2{ q}}{(2\pi)^2}\frac{4D^2\times q^4}{\Big[\Omega^2+(\nu_xq_x^2+\nu_yq_y^2+2\overline\nu q_xq_y)^2\Big]^2},
\end{align}
where we set $k=0$, $\omega =0$. After performing the $\Omega$ integral, above integral reduces to
\begin{align}
&\frac{\lambda^2 D^2}{4} \int\frac{d^2q}{(2\pi)^2}\frac{q^4}{(\nu_x q_x^2 + 2\overline{\nu} q_x q_y + \nu_y q_y^2)^3}, \\
&= \frac{\lambda^2 D^2}{4}(I_4+2I_5+I_8), \\
&=\frac{\lambda^2 D^2}{4}\frac{3}{16 \nu_x^3 p^5}\frac{\ln b}{2\pi}\Big[\beta^2 + \frac{2(\beta+2\alpha^2)}{3}+1\Big].\label{D_corr}
\end{align}

\subsection{One-loop corrections to $\nu_{ij}$}

The one-loop Feynman diagram, with a symmetry factor 8, that contributes to the fluctuation correction of $\nu_x$, $\nu_y$, and $\overline \nu$ is shown in Fig.~\ref{nu_diag}.
\begin{figure}[b]
\centering
\includegraphics[width=0.4\textwidth]{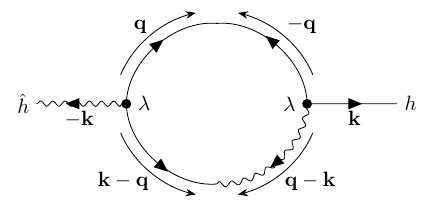}
\caption{One-loop Feynman diagram that contributes to the correction of $\nu_x$, $\nu_y$, and $\overline \nu$.}
\label{nu_diag}
\end{figure}
It is given by
\begin{align*}
&-2D\lambda^2\int_{{\bf q},\Omega}\frac{1}{\big[\Omega^2+(\nu_xq_x^2+\nu_yq_y^2+2\overline\nu q_xq_y)^2\big]}\\ &\times \frac{q_i(k-q)_iq_jk_j}{\big[i\Omega + \nu_x(k-q)_x^2+\nu_y(k-q)_y^2+2\overline\nu (k-q)_x(k-q)_y\big]}.
\end{align*}  
After symmetrizing, i.e., ${\bf q}\rightarrow {{\bf q}+{{\bf k}}/{2}}$, and performing the $\Omega$ integral, we are left with two integrals that include corrections for $\nu_x$, $\nu_y$ and $\overline \nu$. These integrals are
\begin{align}
\mathcal {I}_1=(k_x^2+k_y^2)\frac{\lambda^2 D}{16\nu_x^2p^3}(1+\beta)\int_{\Lambda/b}^{\Lambda}\frac{dq_y}{2\pi q_y},\label{eqn_1}
\end{align}
and
\begin{align}
&\mathcal {I}_2=-2D\lambda^2\int_{q_x,q_y}\frac{(q_xk_x+q_yk_y)q^2}{4(\nu_xq_x^2+\nu_yq_y^2+2\overline\nu q_xq_y)^3}\nonumber \\&\quad\quad\quad\quad\times\big[k_x(\nu_xq_x+\overline\nu q_y)+k_y(\nu_yq_y+\overline\nu q_x)\big].\label{eqn_2}
\end{align} 
Corrections to $\nu_x$ ($\nu_y$) are given by the coefficient of $k_x^2$ ($k_y^2$). Considering the coefficient of $k_x^2$, the correction to $\nu_x$ coming from $\mathcal {I}_2$ is
\begin{align}
&-k_x^2\frac{\lambda^2D}{2}\int_{q_x,q_y}\frac{\nu_xq_x^4+\nu_xq_x^2q_y^2+\overline\nu q_x^3q_y+\overline\nu q_y^3q_x}{(\nu_xq_x^2+\nu_yq_y^2+2\overline\nu q_xq_y)^3},\\
&=-k_x^2\frac{\lambda^2D}{2}\big[\nu_xI_4+\nu_xI_5+\overline\nu I_6 +\overline\nu I_7\big],\\
&=-k_x^2\frac{\lambda^2D}{2}\frac{3}{16\nu_x^2p^3}\bigl[\beta+1/3\bigl]\int_{\Lambda/b}^\Lambda\frac{dq_y}{2\pi q_y}.
\end{align}
Thus collecting coefficient of $k_x^2$ from $\mathcal {I}_1$ and adding it with above equation we obtain correction to $\nu_x$,
\begin{align}
\frac{\lambda^2D}{16\nu_x^2p^3}\frac{\ln b}{2\pi}\bigg[\frac{1}{2}-\frac{\beta}{2}\bigg].\label{nu_x_corr}
\end{align}
To compute the correction to $\nu_y$, we extract the coefficient of $k_y^2$ from integral $\mathcal{I}_2$,
\begin{align}
&-k_y^2(2D\lambda^2)\int_{q_x,q_y}\frac{(q_x^2q_y+q_y^3)(\nu_yq_y+\overline\nu q_x)}{4(\nu_xq_x^2+\nu_yq_y^2+2\overline\nu q_xq_y)^3},\\
&=-k_y^2\frac{\lambda^2D}{2}\big[\nu_yI_5+\nu_yI_8+\overline\nu I_6 +\overline\nu I_7\big],\\
&=-k_y^2\frac{\lambda^2D}{16\nu_x^2p^3}\bigl[3/2+\beta/2\bigl]\int_{\Lambda/b}^\Lambda\frac{dq_y}{2\pi q_y}.
\end{align}
Again collecting coefficient of $k_y^2$ from $\mathcal {I}_1$ and adding it with above equation we obtain correction to $\nu_y$,
\begin{align}
\frac{\lambda^2D}{16\nu_x^2p^3}\frac{\ln b}{2\pi}\bigg[\frac{\beta}{2}-\frac{1}{2}\bigg].\label{nu_y_corr}
\end{align}
Next we collect coefficient of $2k_xk_y$ from $\mathcal {I}_2$ to obtain correction to $\overline \nu$ which is,
\begin{align}
&-2k_xk_y\lambda^2D\int_{q_x,q_y}\frac{q^2q_x(\nu_yq_y+\overline\nu q_x)+q^2q_y(\nu_xq_x+\overline\nu q_y)}{4(\nu_xq_x^2+\nu_yq_y^2+2\overline\nu q_xq_y)^3},\\
&=-(2k_xk_y)\frac{\lambda^2D}{4}\big[(\nu_x+\nu_y)(I_6+I_7)+\overline\nu (I_4+I_8+2I_5)\big].
\end{align}
After some algebra, and collecting coefficient of $2k_xk_y$ from above equation we obtain the correction to $\overline \nu$ which is
$\frac{\lambda^2D\alpha}{16\nu_x^2p^3}\frac{\ln b}{2\pi}.$

\subsection{Nonrenormalization of $\lambda$}
\begin{figure}[h]
\centering
\includegraphics[width=0.3\textwidth]{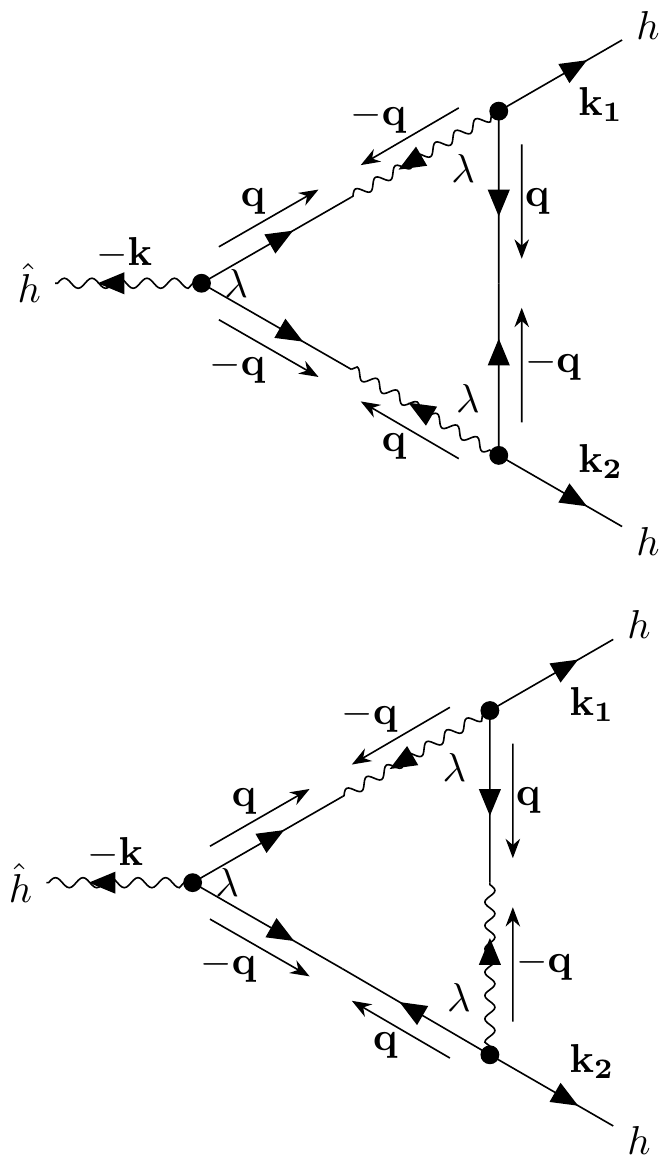}\hspace{0.5in}
\caption{One-loop Feynman diagrams giving the  corrections to  $\lambda$.}
\label{lam_diag}
\end{figure}
The upper diagram in Fig.~\ref{lam_diag} has a symmetry factor 24 and includes corrections to $\lambda$. We find
\begin{align}
&-\lambda^3 D\int_{{\bf q},\Omega}\frac{q^2(-k_{1m}q_m)(q_nk_{2n})}{(-i\Omega+\nu_x q_x^2+2\overline\nu q_xq_y+\nu_y q_y^2)}\times\nonumber \\ &\frac{1}{(i\Omega+\nu_x q_x^2+2\overline\nu q_xq_y+\nu_y q_y^2)\Bigl(\Omega^2+(\nu_x q_x^2+2\overline\nu q_xq_y+\nu_y q_y^2)^2\Bigl)}, \nonumber \\ &=\frac{\lambda^3D}{4}\int_{\bf q}\frac{q^2(k_{1m}q_m)(q_nk_{2n})}{(\nu_x q_x^2+2\overline\nu q_xq_y+\nu_y q_y^2)^3}.\label{lam_corr_1}
\end{align}
Similarly the lower diagram in Fig.~\ref{lam_diag} has a symmetry factor $48$ and contributes 
\begin{align}
&-\lambda^3(2D)\int_{{\bf q},\Omega}\frac{q^2(-k_{1m}q_m)(-q_nk_{2n})}{(-i\Omega+\nu_x q_x^2+2\overline\nu q_xq_y+\nu_y q_y^2)^2}\nonumber \\ &\gap\times\frac{1}{\Bigl[\Omega^2+(\nu_x q_x^2+2\overline\nu q_xq_y+\nu_y q_y^2)^2\Bigl]},\nonumber \\ &=-\frac{\lambda^3D}{4}\int_{\bf q}\frac{q^2(k_{1m}q_m)(q_nk_{2n})}{(\nu_x q_x^2+2\overline\nu q_xq_y+\nu_y q_y^2)^3}.\label{lam_corr_2}
\end{align}   
Equations~(\ref{lam_corr_1}) and (\ref{lam_corr_2}), when added, give zero contribution to the correction of $\lambda$. These are consistent with the Galilean invariance of Eq.~(\ref{aniso-kpz}).
}
\label{app:D}
\subsection{RG recursion relations}

We integrated out the dynamical fields with wave-vectors $k_y$ in the range ($\Lambda/b \leq k_y \leq \Lambda$ and $-\Lambda \leq k_y \leq -\Lambda/b$) in the action functional (\ref{action}). Now we rescale space, time, and fields to raise the  wave-vector cutoff of $k_y$ to { $\pm\Lambda$}. We rescale
 ${\bf k} \to {\bf k}^{'}=b{\bf k} \implies {\bf x} \to {\bf x}^{'}=\frac{{\bf x}}{b}$; similarly, $\omega \to \omega^{'}=b^z\omega \implies t{'}=\frac{t}{b^z}$. Furthermore, the fields are rescaled as 
\begin{align*}
    &h^<({\bf k},\omega)=\xi h({\bf k}^{'},\omega^{'}),\\
     &\hat h^<({\bf k},\omega)=\hat \xi \hat h({\bf k}^{'},\omega^{'}).
\end{align*}
Using these and imposing the condition that coefficient of $\int_{{\bf x},t}\hat{h}\partial_th$ remains unity under rescaling we find that, model parameters scale as
\begin{align*}
&\nu_{ij}^{'}=\nu_{ij}^<b^{z-2},\\
&D^{'}=D^<b^{d+3z}\xi^{-2},\\
   &\lambda^{'}=\lambda^< b^{-d-2}\xi.
\end{align*}
Also, $h^<({\bf x},t)$ scale as $b^{-d-z}\xi h({\bf x}^{'},t^{'})$. Defining $\xi_R=b^{-d-z}\xi=b^\chi$ we get model parameters rescale as
\begin{align*}
&D^{'}=D^<b^{z-d-2\chi},\\
   &\nu_x^{'}=\nu_x^<b^{z-2},\\
   &\nu_y^{'}=\nu_y^<b^{z-2},\\
   &\overline\nu^{'}=\overline\nu^<b^{z-2},\\
   &\lambda^{'}=\lambda^<b^{ \chi+z-2}.
\end{align*}
Here $\chi$ is the roughness exponent of an anisotropic KPZ surface. Since there is no correction to $\lambda$, so $\lambda^<=\lambda$. We further get
\begin{widetext}
\begin{eqnarray}
D^{<}&=&D \biggr[ 1+\frac{2\lambda^2D}{16(2\pi)(\nu_x\nu_y-\overline\nu^2)^{\frac{3}{2}}}\Bigr(\frac{3\beta^2}{4}+\frac{1}{2}(\beta+2\alpha^2)+\frac{3}{4}\Bigr)\frac{\ln b}{(\beta-\alpha^2)}\biggr],\\
\nu_x^{<}&=&\nu_x \biggr[ 1-\frac{2\lambda^2D}{16(2\pi)(\nu_x\nu_y-\overline\nu^2)^{\frac{3}{2}}}\Bigr(\frac{1}{2}-\frac{\beta}{2}\Bigr)\ln b\biggr],\\
\nu_y^{<}&=&\nu_y \biggr[ 1-\frac{2\lambda^2D}{16(2\pi)(\nu_x\nu_y-\overline\nu^2)^{\frac{3}{2}}}\Bigr(\frac{1}{2}-\frac{1}{2\beta}\Bigr)\ln b\biggr],\\
\overline\nu^{<}&=&\overline\nu \biggr[ 1-\frac{2\lambda^2D}{16(2\pi)(\nu_x\nu_y-\overline\nu^2)^{\frac{3}{2}}}\ln b\biggr].
\end{eqnarray}
\end{widetext}
{The extra factor of $2$ is multiplied in numerator of the above renormalized parameters to include contribution from negative values of $q_y$ i.e., ($-\Lambda \leq q_y \leq -\Lambda/b$) in the loop integrals. See the lower strip in Fig.~\ref{int_diag}}. Now set $b=e^{\delta l}\approx 1+\delta l$ for small $\delta l$, and define a dimensionless coupling constant by $g=\frac{\lambda^2D}{16\pi(\nu_x\nu_y-\overline\nu^2)^{\frac{3}{2}}}$ and dimensionless ratios $\alpha = \frac{\overline\nu}{\nu_x}$ and $\beta = \frac{\nu_y}{\nu_x}$ (see main text). These allow us to obtain the differential RG recursion relations reported in the main text.

\section{Connections to other anisotropic KPZ equations}

We begin with the most general anisotropic KPZ equation
{
\begin{align}
&\frac{\partial h}{\partial t}
+ \frac{\lambda_x'}{2} \Big(\frac{\partial h}{\partial x}\Big)^2
+ \frac{\lambda_y'}{2} \Big(\frac{\partial h}{\partial y}\Big)^2
+ \overline{\lambda}' \, \frac{\partial h}{\partial x} \, \frac{\partial h}{\partial y} \nonumber \\
& \quad\quad\quad = \nu_x' \frac{\partial^2 h}{\partial x^2}
+ \nu_y' \frac{\partial^2 h}{\partial y^2}
+ 2 \overline{\nu}' \frac{\partial^2 h}{\partial x \partial y}
+ f .\label{an-kpz}
\end{align}}
 The nonlinear terms in (\ref{an-kpz}) may be written in a compact form as $\lambda_{ij}(\partial_ih)(\partial_jh)/2$,
where the matrix 

\begin{equation}
\lambda_{ij} \equiv
\begin{pmatrix}
\lambda_x' & \overline{\lambda}' \\
\overline{\lambda}' & \lambda_y'
\end{pmatrix}
\end{equation}

\noindent is a symmetric matrix and hence can be diagonalized by an appropriate similarity transformation [which here is a coordinate transformation, i.e., { ($x'=x \cos \theta + y \sin \theta$ and $y'=-x \sin \theta + y \cos \theta$)}, defined by a rotation angle $\theta$]. The rotation angle $\theta$ is obtained by requiring that the coefficient of mixed derivative term $\frac{\partial h}{\partial x} \, \frac{\partial h}{\partial y}$ vanishes after rotation, which leads to

\begin{equation}
\tan 2\theta = \frac{2\overline{\lambda}'}{\lambda_x'-\lambda_y'}.
\end{equation}

In this rotated frame, the equation of motion reduces to
{
\begin{align}
&\frac{\partial h}{\partial t}
+ \frac{\lambda_x}{2} \Big(\frac{\partial h}{\partial x}\Big)^2
+ \frac{\lambda_y}{2} \Big(\frac{\partial h}{\partial y}\Big)^2 \nonumber \\
& = \nu_x \frac{\partial^2 h}{\partial x^2}
+ \nu_y \frac{\partial^2 h}{\partial y^2}
+ 2 \overline{\nu} \frac{\partial^2 h}{\partial x \partial y}
+ f ,
\label{full-akpz}
\end{align}}
where the linear couplings remain, in general, anisotropic and may include an off-diagonal term. Here
\begin{eqnarray}
\lambda_x &=& \lambda_x'\cos^2\theta + \lambda_y'\sin^2\theta+\overline\lambda'\sin 2\theta, \\
\lambda_y &=& \lambda_x'\sin^2\theta + \lambda_y'\cos^2\theta-\overline\lambda'\sin 2\theta, \\
\nu_x &=& \nu_x'\cos^2\theta + \nu_y'\sin^2\theta+\overline\nu'\sin 2\theta, \\
\nu_y &=& \nu_x'\sin^2\theta + \nu_y'\cos^2\theta-\overline\nu'\sin 2\theta, \\
\overline\nu &=& -\nu_x'\sin\theta\cos\theta+\nu_y'\sin\theta\cos\theta+\overline\nu'\cos 2\theta,
\end{eqnarray}
are the modified linear and nonlinear couplings.
We consider the case in which the nonlinear couplings $\lambda_x$ and $\lambda_y$ have the same sign. In this situation, we perform an anisotropic rescaling of the spatial coordinates as
\begin{equation}
x \rightarrow x' = \sqrt{\frac{\lambda_y}{\lambda_x}} \, x,
\qquad
y \rightarrow y' = y,
\end{equation}
which
brings Eq.~(\ref{full-akpz}) to the form
\begin{align}
{\frac{\partial h}{\partial t}
+ \frac{\lambda_y}{2} (\boldsymbol{\nabla} h)^2
= \tilde\nu_x \frac{\partial^2 h}{\partial x^2}
+ \nu_y \frac{\partial^2 h}{\partial y^2}
+ 2 \tilde{\nu} \frac{\partial^2 h}{\partial x \partial y} + f, \label{reduced_akpz}}
\end{align}
with transformed coefficients
\begin{equation}
\tilde\nu_x = \frac{\lambda_y}{\lambda_x} \nu_x,
\qquad
\tilde{\nu} = \sqrt{\frac{\lambda_y}{\lambda_x}} \, \overline{\nu}.
\end{equation}

{Equation~(\ref{reduced_akpz}) is exactly the same as Eq.~(\ref{aniso-kpz}), with the identification ($\lambda_y,\tilde\nu_x,\nu_y,\tilde\nu) \equiv (\lambda,\nu_x,\nu_y,\overline\nu$), as discussed in the main text, which we have used to describe an active anisotropic fluid membrane.} Next, notice that the linear surface tension tensor appearing in Eq.~(\ref{reduced_akpz}) is
symmetric and can therefore be brought to a diagonal form by a rotation of the spatial coordinates  using the same transformation introduced earlier. The rotation angle $\tilde\theta$ is
fixed by the condition
\begin{equation}
\tan 2\tilde\theta = \frac{2\tilde{\nu}}{\tilde{\nu}_x - \nu_y}.\label{theta_eq}
\end{equation}
After performing this transformation, the mixed second derivative term is
eliminated and Eq.~(\ref{reduced_akpz}) takes the form
{\begin{eqnarray}
\frac{\partial h}{\partial t}
+ \frac{\lambda_y}{2}(\boldsymbol{\nabla} h)^2
&=&
\nu_{xx}\frac{\partial^2 h}{\partial x^2}
+
\nu_{yy}\frac{\partial^2 h}{\partial y^2}
+ f,
\label{wolf_akpz}
\end{eqnarray}}
with the diagonal coefficients given by
\begin{eqnarray}
\nu_{xx} &=& \tilde{\nu}_x \cos^2\tilde\theta
+ \nu_y \sin^2\tilde\theta
+ \tilde{\nu}\sin 2\tilde\theta,\label{nu_xx_eq} \\
\nu_{yy} &=& \tilde{\nu}_x \sin^2\tilde\theta
+ \nu_y \cos^2\tilde\theta
- \tilde{\nu}\sin 2\tilde\theta.\label{nu_yy_eq}
\end{eqnarray}
Equation~(\ref{wolf_akpz}) is same as the anisotropic KPZ model studied in the earlier work \cite{wolf,frey-tauber}.
 
{ 
Let us now define an isotropic damping coefficient $\nu_{\mathrm{iso}}$ and an anisotropy parameter $\nu_{\mathrm{aniso}}$ as
\begin{eqnarray}
\nu_{\mathrm{iso}} &=& \frac{\nu_{xx}+\nu_{yy}}{2}
= \frac{\tilde{\nu}_{x}+\nu_{y}}{2},\\[4pt]
\nu_{\mathrm{aniso}} &=& \nu_{xx}-\nu_{yy}
= \sqrt{(\tilde{\nu}_x-\nu_y)^2+4\tilde{\nu}^{\,2}}.
\end{eqnarray}
Although the three coefficients $(\tilde{\nu}_x,\,\nu_y,\,\tilde{\nu})$ can be equivalently reparametrized in terms of $(\nu_{\mathrm{iso}},\,\nu_{\mathrm{aniso}},\,\tilde{\theta})$, where $\tilde{\theta}$ denotes the rotation angle specifying the orientation of the principal axes of the diffusion tensor, the physical anisotropy is fully characterized by the eigenvalue difference $\nu_{\mathrm{aniso}}$, which depends on $\tilde{\nu}_x$, $\nu_y$, and $\tilde{\nu}$.\\

\begin{figure*}[]
 \includegraphics[width=\textwidth]{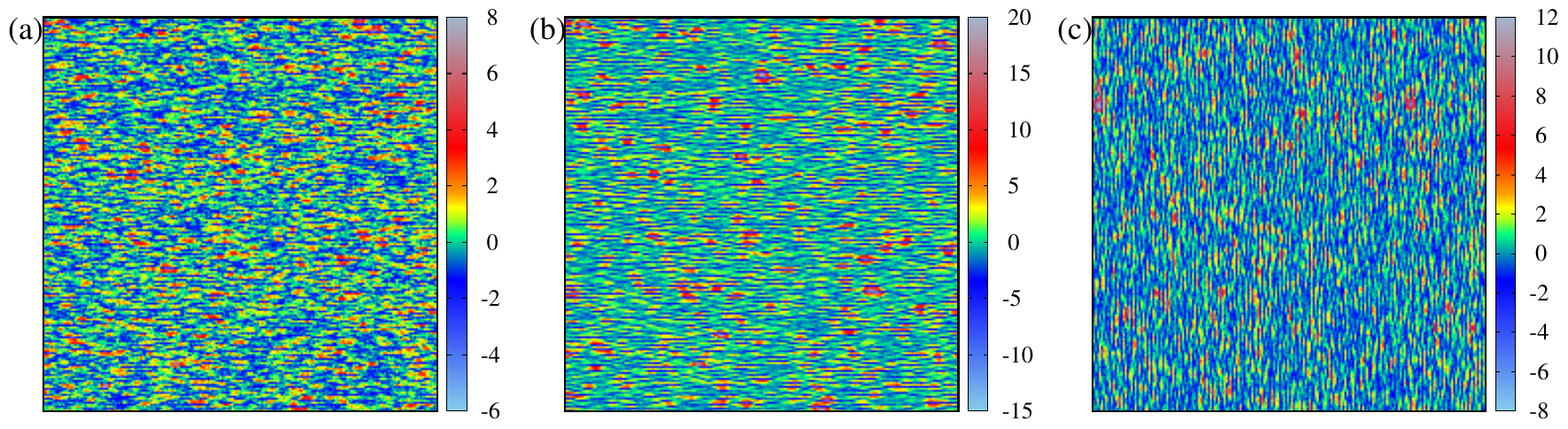}
\caption{Snapshots of the height profiles (\(L=200\)) in the linearly unstable regime before the divergence of the interface fluctuations, with parameters, (a) \(\nu_{xx}=1.68\), \(\nu_{yy}=-0.48\), \(\nu_4=0.10\), \(D=1.0\), and \(\lambda=2.20\); (b) \(\nu_{xx}=3.0\), \(\nu_{yy}=-1.0\), \(\nu_4=0.10\), \(D=1.0\), and \(\lambda=2.30\). In both (a) and (b) ripples form parallel to the \(x\) axis. (c) Same parameters as (b), except \(\nu_{xx}=-1.0\) and \(\nu_{yy}=3.0\), resulting in ripples parallel to the \(y\) axis.
}\label{crumple_plot_2}
\end{figure*}

Considering only the linear part of Eq.~(\ref{wolf_akpz}), the evolution equation in Fourier space becomes
\begin{align}
\partial_t h = -\nu_{xx}k^2\left(\cos^2\theta+\beta\sin^2\theta\right)h,
\end{align}
where $k_x=k\cos\theta$, $k_y=k\sin\theta$, and $\beta=\nu_{yy}/\nu_{xx}$. For linear stability, the coefficient multiplying $h$ must remain positive for all orientations of $\theta$, i.e.,
\begin{align}
1+(\beta-1)\sin^2\theta >0.
\end{align}
When $\nu_{xx}>0$ and $\nu_{yy}<0$, we have $\beta<0$, and the above condition cannot be satisfied for all $\theta$. Consequently, the interface is linearly unstable. The instability grows most rapidly along the direction for which
\begin{align}
1+(\beta-1)\sin^2\theta
\end{align}
attains its most negative value. This occurs at $\theta=\pm\pi/2$, corresponding to wave vectors oriented along the $y$ direction. Following the analysis presented in Appendix~\ref{app:B}, one therefore expects the emergence of stripes parallel to the $x$ axis prior to the divergence of the interface fluctuations.

Conversely, when $\nu_{yy}>0$ and $\nu_{xx}<0$, the most unstable mode is oriented along the $x$ direction, leading to the formation of stripes parallel to the $y$ axis before the instability develops fully (see Appendix~\ref{app:F}).
}

Notice that under renormalization group transformations on Eq.~(\ref{aniso-kpz}) of the main text, the off-diagonal linear term $\overline{\nu}$ turns out to be irrelevant (in the RG sense) in the long-wavelength limit and flows to zero. Consequently, the effective large-scale theory reduces to the anisotropic KPZ model previously studied in Refs.~\cite{wolf,frey-tauber}. Thus our RG results are consistent with the symmetry arguments elucidated above. Furthermore, when $\nu_x \to \nu_y$, the system becomes effectively isotropic and crosses over to the standard KPZ universality class.

Importantly, Eq.~(\ref{full-akpz}) cannot be mapped onto the previously studied anisotropic KPZ models by any coordinate transformation when $\lambda_x$ and $\lambda_y$ have opposite signs, a situation not directly relevant for an active membrane (and hence is not studied here).\label{app:E}

\section{Direct numerical simulations}

Following Ref.~\cite{wolf_num}, we discretize Eq.~(\ref{aniso-kpz}) by standard forward-backward differences on a square grid with lattice constant $a_0$. Employing the Euler algorithm with time increments $\Delta t$, we integrate Eq.~(\ref{aniso-kpz}). The grid points are labeled by ${\bf n}$, and the basis vector is denoted by ${\bf e}_i$, where $i=1,2$, giving the discretized form of Eq.~(\ref{aniso-kpz})

\begin{widetext}
{
\begin{align} 
h_{\bf n}(t+\Delta t)&=h_{\bf n}(t)+\frac{\Delta t}{(a_0)^2}\sum_{i=1}^2\biggl\{ \nu_i \Bigl[h_{\bf n+e_i}(t)-2h_{\bf n}(t)+h_{\bf n-e_i}(t)\Bigl]-\frac{\lambda}{8}\Bigl[h_{\bf n+e_i}(t)-h_{\bf n-e_i}(t)\Bigl]^2\biggl\}+\frac{\overline \nu \Delta t}{2 (a_0)^2}\Bigl[h_{\bf n+e_1+e_2}(t)\nonumber \\ &-h_{\bf n+e_1-e_2}(t)+h_{\bf n-e_1-e_2}(t)-h_{\bf n-e_1+e_2}(t)\Bigl]+\sqrt{\frac{24 D \Delta t}{(a_0)^2}}R_{\bf n}(t).\label{des_akpz_2}
\end{align}
}
\end{widetext}
Here $\nu_1(\nu_2) \equiv \nu_x(\nu_y)$. The variable $R$ corresponds to uniformly distributed random numbers ranging from $-\frac{1}{2}$ to $\frac{1}{2}$. The prefactor $\sqrt{\frac{24 D \Delta t}{(a_0)^2}}$ ensures that the noise possesses an equivalent second moment as Gaussian noise integrated over a time interval $\Delta t$.

For our simulations we chose $a_0=1$.  In the the speculated algebraic rough phase region, width ${\cal W}$ is expected to scale as $\mathcal{W}(L,t)\sim L^{\chi}f(\frac{t}{L^z})$, where $f(x)$ is a dimensionless scaling function.
With different set of parameters we run the simulation for $L=100$, $200$, $300$, $400$, and $500$. For both of the growth and saturation region ${\cal W}$ was calculated at different times with $N$ noise averages up to time $T$. For system size $L=100$, $200$, $300$, $400$, and $500$ Eq.~(\ref{des_akpz_2}) was simulated up to time $T=5000$, $10000$, $15000$, $25000$, and $30000$ with time step $\Delta t=0.01$. $L=100$, $200$ was noise averaged with $N=500$. For $L=300$, $400$, and $500$ $N=200$, $100$, and $50$ was taken. Refer to Fig.~\ref{num_sat_region} in the main text for the variation of ${\cal W}$ over time on a log-log scale. With $T$ defined earlier for different $L$, we reached the saturation region of ${\cal W}$. Then we numerically solved the equation of motion (\ref{aniso-kpz}) for further $100000$ time steps with $\Delta t=0.01$. And for each of these $100000$ time steps we have taken $N=500$ noise averages for all $L=100$ to $500$. For each of these time steps ${\cal W}$ was calculated and averaged over $N$. The final ${\mathcal W_\text{sat}}$ was calculated by averaging over 100000 realizations in the saturated regime. Variation of ${\cal W}_{\text{sat}}$ with $L$ in the saturation region is shown in  Fig.~\ref{num_sat_region}(d) of the main text.

{Last, we revisit the linearly unstable cases discussed in the main text. The corresponding height snapshots are shown in Figs.~\ref{density_plot}(c) and~\ref{density_plot}(d), where the ripples appear at oblique angles. To understand their orientation, we rotate the coordinate system so that it aligns with the principal axes of the diffusion tensor $\nu_{ij}$. In this basis, the diffusion matrix becomes diagonal, i.e.,
\[
(\nu_x,\nu_y,\overline{\nu}) \rightarrow (\nu_{xx},\nu_{yy}).
\]

For Fig.~\ref{density_plot}(c), the parameters $\nu_x=1.0$, $\nu_y=0.20$, and $\overline{\nu}=1.0$ transform to $\nu_{xx}=1.68$ and $\nu_{yy}=-0.48$. Similarly, for Fig.~\ref{density_plot}(d), the parameters $\nu_x=1.0$, $\nu_y=1.0$, and $\overline{\nu}=2.0$ transform to $\nu_{xx}=3.0$ and $\nu_{yy}=-1.0$. In both cases, $\nu_{yy}<0$, rendering the system linearly unstable along the $y$ direction. Consequently, the fastest growing mode has its wave-vector oriented along the $y$ axis, leading to the formation of ripples parallel to the $x$ axis, as shown in Figs.~\ref{crumple_plot_2}(a) and~\ref{crumple_plot_2}(b).

For completeness, Fig.~\ref{crumple_plot_2}(c) illustrates the complementary situation in which $\nu_{xx}<0$ and $\nu_{yy}>0$. In this case, the instability develops along the $x$ direction, resulting in ripples oriented parallel to the $y$ axis.
}

\label{app:F}

\bibliography{anisokpz.bib}

\begin{thebibliography}{67}
\bibitem [{sup()}]{supple}%
  \BibitemOpen
  \bibfield  {title} {\enquote {\bibinfo {title} {{T}he corresponding movies
  {MOV}1 [for {F}ig.~\ref{density_plot}(a)], {MOV}2 [for
  {F}ig.~\ref{density_plot}(c)] and {MOV}3 [for {F}ig.~\ref{density_plot}(d)]
  are available in the {S}upplemental {M}aterial ({SM}).}}\ }\href@noop {} {\
  }\BibitemShut {NoStop}%
\end{thebibliography}
\end{document}